\documentstyle[12pt]{article}

\def\o{\over}
\def\b{\begin{equation}}
\def\e{\end{equation}}
\def\l{\label}

\def\kpmm{$K^+\rightarrow\pi^+\mu^+\mu^-$\ }

\def\imlt{{\rm Im}\lambda_t}
\def\relt{{\rm Re}\lambda_t}
\def\relc{{\rm Re}\lambda_c}

\def\a{{\alpha_s\o 4\pi}}

\def\r#1{(\ref{#1})}

\begin{document}
\thispagestyle{empty}
\begin{flushright}
 MPI-PhT/94-43 \\
 TUM-T31-65/94 \\
 June 1994
\end{flushright}
\vskip1truecm
\centerline{\Large\bf Parity Violating Longitudinal Muon Polarization}
\centerline{\Large\bf in $K^+\to\pi^+\mu^+\mu^-$}
\centerline{\Large\bf Beyond Leading Logarithms
   \footnote[1]{\noindent
   Supported by the German
   Bundesministerium f\"ur Forschung und Technologie under contract
   06 TM 732 and by the CEC science project SC1--CT91--0729.}}
\vskip1truecm
\centerline{\sc Gerhard Buchalla {\rm and} Andrzej J. Buras}
\bigskip
\centerline{\sl Max-Planck-Institut f\"ur Physik}
\centerline{\sl  -- Werner-Heisenberg-Institut --}
\centerline{\sl F\"ohringer Ring 6, D-80805 M\"unchen, Germany}
\vskip0.6truecm
\centerline{\sl Technische Universit\"at M\"unchen, Physik Department}
\centerline{\sl D-85748 Garching, Germany}

\vskip1truecm
\centerline{\bf Abstract}
We generalize the existing analyses of the parity violating
muon polarization asymmetry
$\Delta_{LR}$ in $K^+\to\pi^+\mu^+\mu^-$ beyond the leading logarithmic
approximation. The inclusion of next--to--leading QCD corrections reduces
the residual dependence on the renormalization scales, which is quite
pronounced in the leading order. This leads to a considerably
improved accuracy in the perturbative calculation of the
short distance dominated quantity $\Delta_{LR}$.
Accordingly this will also allow to obtain better constraints on
the Wolfenstein parameter $\varrho$ from future measurements
of $\Delta_{LR}$.
For
$-0.25\leq \varrho \leq 0.25$, $V_{cb}=0.040\pm 0.004$ and
$m_t=(170\pm 20)GeV$ we find
$3.0\cdot 10^{-3}\leq |\Delta_{LR}|\leq 9.6\cdot 10^{-3}$.
\vfill
\newpage

\pagenumbering{arabic}

It has been pointed out by Savage and Wise \cite{SW1} that
measurements of muon
polarization in $K^+\to\pi^+\mu^+\mu^-$ decay can give valuable
information on the weak mixing angles and in particular on the parameter
$\varrho$ in the Wolfenstein parametrization \cite{WO}. Indeed as shown in
\cite{SW1,SW2},
the parity-violating asymmetry
\begin{equation}\label{1}
\mid\Delta_{LR}\mid=\mid\frac{\Gamma_R-\Gamma_L}{\Gamma_R+\Gamma_L}\mid
= r \cdot\mid {\rm Re}\xi \mid
\end{equation}
is dominated by the short distance contributions of Z-penguin
and W-box dia\-grams with internal charm and top quark exchanges,
while the total rate is completely determined by the one-photon
exchange amplitude. The interference of this leading amplitude with
the small short distance piece is the source of the asymmetry
$\Delta_{LR}$.
Here $\Gamma_R$ ($\Gamma_L$) is the rate to produce right-
(left-) handed $\mu^+$, that is $\mu^+$ with spin along (opposite to)
its three-momentum direction.
The factor $r$ arises from phase space integrations. It depends
only on the particle masses $m_K$, $m_\pi$ and $m_\mu$, on the
form factors of the matrix element
$\langle\pi^+\mid(\bar sd)_{V-A}\mid K^+\rangle$,
as well as on the form factor of the $K^+\to\pi^+\gamma^\ast$
transition, relevant for the one-photon amplitude. In addition $r$
depends on a possible cut which may be imposed on $\theta$, the
angle between the three-momenta of the $\mu^-$ and the pion in the rest
frame of the $\mu^+\mu^-$ pair.
Without any cuts one has
$r=2.3$ \cite{SW2}. If $\cos\theta$ is restricted to lie in the region
$-0.5\leq\cos\theta\leq 1.0 $, this factor is increased to $r=4.1$.
As discussed in \cite{SW2}, such a cut in $\cos\theta$ could be
useful since it enhances $\Delta_{LR}$ by 80\% with only a 22\%
decrease in the total number of events.
\\
${\rm Re}\xi$ is a purely short distance function depending only
on CKM parameters, the QCD scale
 $\Lambda \overline{_{MS}}$
 and the
quark masses $m_t$ and $m_c$. We will discuss it in detail below.

$\Delta_{LR}$ as given in (\ref{1}) has also been considered by
B\'elanger et al. \cite{GT1}, who emphasized its close relation to the
short distance part of the decay amplitude $K_L\to\mu^+\mu^-$.
Unfortunately the authors of ref. \cite{GT1}
did not include the internal charm
contribution to $\Delta_{LR}$. As we will show explicitly below the
charm contribution cannot be neglected as its presence increases the
extracted value of $\varrho$ by roughly $\Delta\varrho=0.2$.

Let us briefly summarize the theoretical situation of $\Delta_{LR}$.
The "kinematical" factor $r$ can be essentially obtained from
experimental input on the particle masses and the form factors.
The form factor $f$ describing the $K^+\to\pi^+\gamma^\ast$ vertex
has been discussed in detail in \cite{EPR} within chiral
perturbation theory. While the imaginary part can be reliably
predicted, the real part is only determined up to a constant to
be fitted from experiment. On the other hand, data on
$K^+\to\pi^+e^+e^-$ decay \cite{ALL} allow to extract the absolute
value of this form factor directly. We will adopt this approach,
following \cite{SW2}. Since the imaginary part of $f$ is quite
small \cite{SW2,EPR}, we then also have the real part ${\rm Re}f$.
In principle ${\rm Im}f$ could yield an extra contribution in
\r{1} proportional to ${\rm Im}\xi$. We have checked, based on the
approach of \cite{SW2} that this contribution is below 1\% of the
dominant part shown in \r{1} and can therefore be safely
neglected. Clearly the factor $r$ involves some uncertainty due
to the experimental errors in the form factors, which can however
be further reduced by future improved measurements. For the
present discussion we will assume fixed numerical values for $r$.
\\
Besides the short distance part of $\Delta_{LR}$ there are also
potential long distance contributions coming from two-photon
exchanges, which have also been discussed by the authors of \cite{SW2}.
These are difficult to calculate in a reliable manner, but the
estimates given in \cite{SW2} indicate that this contribution
is substantially smaller than the short distance part, although
it cannot be fully neglected.
Therefore the short distance effects are expected to safely
dominate the quantity $\Delta_{LR}$ and we shall concentrate our
discussion on this part, keeping in mind the possible existence
of non-negligible long distance corrections.
\\
The short distance physics leading to $\Delta_{LR}$ is generally
considered to be very clean, as it can be treated within a
perturbative framework. However this does not mean that it is
free of theoretical uncertainties. An indication of the involved
error due to the necessary truncation of the perturbation series
in the strong coupling constant $\alpha_s$ can be obtained by
studying the sensitivity of a physical quantity to the relevant
renormalization scales on which it should not depend in principle.
The existing short distance calculations of
$\Delta_{LR}$
\cite{SW1,SW2,GT1} include QCD
corrections in the leading logarithmic approximation (LLA) \cite{EH}.
As it turns out, they
suffer from sizable theoretical uncertainties due to the residual
scale dependences.
The main purpose of our letter is to extend the analyses of
\cite{SW1,SW2,GT1}
beyond the leading logarithmic approximation thereby reducing considerably
the theoretical uncertainties in question. To this end we will use our
next-to-leading order analysis of $K_L\to\mu^+\mu^-$ presented in
\cite{BB2,BB3}.

Our discussion of the Cabibbo-Kobayashi-Maskawa matrix
will be based on the standard parametrization \cite{PDG},
which can equivalently be rewritten in terms of the Wolfenstein
parameters ($\lambda$, $A$, $\varrho$, $\eta$) through the
definitions \cite{BLO}
\b\l{wop}
s_{12}\equiv\lambda \qquad s_{23}\equiv A \lambda^2 \qquad
s_{13} e^{-i\delta}\equiv A \lambda^3 (\varrho-i \eta)      \e
The unitarity structure of the CKM matrix is conventionally
represented through the unitarity triangle in the complex plane
with coordinates (0, 0), (1, 0) and ($\bar\varrho$, $\bar\eta$)
where
\b\l{ut}
\bar\varrho + i \bar\eta\equiv -{V_{ud}V^\ast_{ub}\o
 V_{cd}V^\ast_{cb}}   \e
To better than 0.1\% accuracy $\bar\varrho=\varrho(1-\lambda^2/2)$
and $\bar\eta=\eta(1-\lambda^2/2)$.

Following \cite{BB2,BB3} it is straightforward to generalize
the expression for ${\rm Re}\xi$
of \cite{SW2} beyond leading logarithms. We find
\begin{equation}\label{2}
{\rm Re}\xi=
\kappa\cdot\left[{\relc\o\lambda}P_0+{\relt\o\lambda^5}Y(x_t)
\right]            \e
\b\l{kap}
\kappa={\lambda^4\o 2\pi\sin^2\Theta_W(1-{\lambda^2\o 2})}
 =1.66\cdot 10^{-3}   \e
Here $\lambda=|V_{us}|=0.22$, $\sin^2\Theta_W=0.23$,
$x_t=m^2_t/M^2_W$ and $\lambda_i=V^\ast_{is}V_{id}$.
The function $Y$, relevant for the top contribution, is given by
\b\l{yy}
Y(x) = Y_0(x) + \a Y_1(x)\e
\b\l{yy0}
Y_0(x) = {x\over 8}\left[{4-x\over 1-x}+{3x\over (1-x)^2}\ln x\right]
\e
and
\begin{eqnarray}\l{yy1}
Y_1(x) = &&{4x + 16 x^2 + 4x^3 \over 3(1-x)^2} -
           {4x - 10x^2-x^3-x^4\over (1-x)^3} \ln x\nonumber\\
         &+&{2x - 14x^2 + x^3 - x^4\over 2(1-x)^3} \ln^2 x
           + {2x + x^3\over (1-x)^2} L_2(1-x)\nonumber\\
         &+&8x {\partial Y_0(x) \over \partial x} \ln x_\mu
\end{eqnarray}
where $x_\mu=\mu^2/M^2_W$ with $\mu=\mu_t={\cal O}(m_t)$ and
\b\l{l2}
L_2(1-x)=\int^x_1dt{\ln t\o 1-t}   \e
The QCD correction $Y_1$ has been calculated in \cite{BB2}.
Next
\b\l{6}
P_0=\frac{Y_{NL}}{\lambda^4}
 \e
where $Y_{NL}$ represents the renormalization group expression for the
charm contribution in next-to-leading logarithmic approximation (NLLA)
calculated in \cite{BB3}. It reads
\b\l{ynl} Y_{NL}=C_{NL}-B^{(-1/2)}_{NL}  \e
where $C_{NL}$ is the Z-penguin part and
$B^{(-1/2)}_{NL}$ is the box contribution, relevant
for the case of final state leptons with weak isospin $T_3=-1/2$.
We have
\begin{eqnarray}\l{cnln}
\lefteqn{C_{NL}={x(m)\o 32}K^{{24\o 25}}_c\left[\left({48\o 7}K_++
 {24\o 11}K_--{696\o 77}K_{33}\right)\left({4\pi\o\alpha_s(\mu)}+
 {15212\o 1875} (1-K^{-1}_c)\right)\right.}\nonumber\\
&&+\left(1-\ln{\mu^2\o m^2}\right)(16K_+-8K_-)-{1176244\o 13125}K_+-
 {2302\o 6875}K_-+{3529184\o 48125}K_{33} \nonumber\\
&&+\left. K\left({56248\o 4375}K_+-{81448\o 6875}K_-+{4563698\o 144375}K_{33}
  \right)\right]
\end{eqnarray}
where
\b\l{kkc} K={\alpha_s(M_W)\o\alpha_s(\mu)}\qquad
  K_c={\alpha_s(\mu)\o\alpha_s(m)}  \e
\b\l{kkn} K_+=K^{{6\o 25}}\qquad K_-=K^{{-12\o 25}}\qquad
          K_{33}=K^{{-1\o 25}}  \e
\begin{eqnarray}\l{bmnln}
\lefteqn{B^{(-1/2)}_{NL}={x(m)\o 4}K^{24\o 25}_c\left[ 3(1-K_2)\left(
 {4\pi\o\alpha_s(\mu)}+{15212\o 1875}(1-K^{-1}_c)\right)\right.}\nonumber\\
&&-\left.\ln{\mu^2\o m^2}-
  {329\o 12}+{15212\o 625}K_2+{30581\o 7500}K K_2
  \right]
\end{eqnarray}
Here $K_2=K^{-1/25}$, $m=m_c$, $x=m^2/M^2_W$. In \r{cnln} -- \r{bmnln}
the two-loop expression has to be used for $\alpha_s(\mu)$ and
$\mu=\mu_c={\cal O}(m_c)$. The explicit $\mu$-dependences in the
next-to-leading order terms \r{yy1}, \r{cnln} and \r{bmnln} cancel the
scale ambiguity of the leading contributions to the considered order in
$\alpha_s$. The consequences of this feature will be discussed
later on.
Numerical values of $P_0$
are given in table 1 where $m_c\equiv\bar m_c(m_c)$.
\begin{table}
\begin{center}
\begin{tabular}{|c|c|c|c|}\hline
&\multicolumn{3}{c|}{$P_0 $}\\ \hline
$\Lambda_{\overline{MS}}$, $m_c$ [$GeV$]
&1.25&1.30&1.35\\ \hline
0.20&0.132&0.141&0.150\\ \hline
0.25&0.135&0.145&0.154\\ \hline
0.30&0.139&0.148&0.158\\ \hline
0.35&0.142&0.152&0.162\\ \hline
\end{tabular}
\end{center}
\centerline{}
{\bf Table 1:} The function $P_0$ for various
$\Lambda_{\overline{MS}}$ and $m_c$.
\end{table}

It is evident from \r{1} and \r{2} that, given $|\Delta_{LR}|$,
one can extract $\relt$:
\b\l{rltdlr}
\relt=-\lambda^5{{|\Delta_{LR}|\o r \kappa}-
  \left(1-{\lambda^2\o 2}\right) P_0\o Y(x_t)}  \e
Furthermore,
using the standard parametrization of the CKM matrix we obtain
from the definition \r{ut}
\b\l{9}
\bar\varrho={\sqrt{1+4 s_{12}c_{12}\relt /s^2_{23}-
(2 s_{12}c_{12}\imlt /s^2_{23})^2}-
1+2 s^2_{12}\o 2 c^2_{23} s^2_{12}}   \e
Up to the very accurate approximations that $V_{cd}V^\ast_{cb}$ is real
(error below 0.1\%) and
$c_{13}=1$ (error less than $10^{-5}$)
\r{9} is an {\it exact\/} relation.
Using the excellent approximation $\imlt=\eta A^2\lambda^5$
\cite{BLO}, we see that a measured value of $\relt$ determines
by means of \r{9} a curve in the ($\varrho$, $\eta$)-plane.
Since the dependence on $\imlt$ is however very small, this curve
will be almost parallel to the $\eta$-axis.
Thus, knowledge of $|\Delta_{LR}|$, hence $\relt$, implies a value
for $\varrho$ (or $\bar\varrho$) almost independently of $\imlt$.
For simplicity we shall
neglect $\imlt$ in \r{9} completely, which introduces a change in
$\bar\varrho$ of at most 0.01. It is evident that the more general
treatment can be easily restored if desired.
Note that the charm sector contributes to $\bar\varrho$ the
non-negligible portion
\b\l{drhc}
\Delta\bar\varrho_{\rm charm}\approx P_0/(A^2 Y(x_t))\approx 0.2 \e

In order to demonstrate briefly the phenomenological
consequences of the next-to-leading order calculation
we consider the following scenario.
We assume that the asymmetry $\Delta_{LR}$ is known to within $\pm 10\%$
\b\l{11}
\Delta_{LR}=(6.0\pm 0.6)\cdot 10^{-3}
 \e
where a cut on $\cos\theta$, $-0.5\leq\cos\theta\leq 1.0$,
is understood.
Next we take ($m_i\equiv\bar m_i(m_i)$)
\b\l{mtcv}
m_t=(170\pm 5)GeV\quad m_c=(1.30\pm 0.05)GeV\quad
V_{cb}=0.040\pm 0.001  \e
\b\l{lams}
\Lambda_{\overline{MS}}=(0.30\pm 0.05)GeV   \e
The errors quoted here seem quite reasonable if one keeps in mind
that it will take at least ten years to achieve the accuracy
assumed in \r{11}. The value of $m_t$ in \r{mtcv} is in the ball
park of the most recent results of the CDF collaboration \cite{CDF}.

In table 2 we have displayed the central value for $\bar\varrho$
as it is extracted from $\Delta_{LR}$ (\r{rltdlr} and \r{9})
in our example, along with
the uncertainties due to the parameters involved.
\begin{table}
\begin{center}
\begin{tabular}{|c||c||c|c|c|c|c|}\hline
&&$\Delta(\Delta_{LR})$&$\Delta(m_t)$
&$\Delta(V_{cb})$&$\Delta(m_c)$&$\Delta(\Lambda_{\overline{MS}})$
\\ \hline
$\bar\varrho$&$-0.06$&$\pm 0.13$&$\pm 0.05$&$\pm 0.06$&$\pm 0.01$&
$\pm 0.00$\\ \hline
\end{tabular}
\end{center}
\centerline{}
{\bf Table 2:}
$\bar\varrho$ determined
from $\Delta_{LR}$ for the scenario described in the text together
with the uncertainties related to various input parameters.
\end{table}
This is intended to indicate the sensitivity of $\bar\varrho$ to
the relevant input. The combined errors due to a simultaneous variation
of several parameters may be obtained to a good approximation by
simply adding the errors from table 2.
It is interesting to compare these numbers
with the renormalization scale ambiguities, which inevitably limit
the accuracy of the short-distance calculation. If we vary the scale
in the charm- and in the top sector as $1GeV\leq\mu_c\leq 3GeV$
and $100GeV\leq\mu_t\leq 300GeV$, respectively, keeping all other
parameters at their central values, we obtain the following range
for $\bar\varrho$
\b\l{rhor}
-0.15\leq\bar\varrho\leq -0.03 \qquad {\rm (NLLA)}   \e
\b\l{rhor0}
-0.31\leq\bar\varrho\leq 0.02 \qquad {\rm (LLA)}   \e
We would like to emphasize the following points:
\begin{itemize}
\item
The error in $\bar\varrho$ from \r{rhor}, which illustrates the
theoretical uncertainty of the short distance piece alone,
is not negligible. It seems however moderate when compared to the
errors shown in table 2. We stress that \r{rhor} is based on the
complete next-to-leading order result for ${\rm Re}\xi$. If only
the leading log approximation is used instead, the range obtained
for $\bar\varrho$ is by almost a factor of 3 larger \r{rhor0}.
\item
The error in \r{rhor} is almost entirely due to the charm sector.
Indeed, if we vary only $\mu_c$, keeping $\mu_t=m_t$ fixed, the
corresponding interval for $\bar\varrho$ reads ($-0.14$, $-0.03$).
This illustrates once more, that the charm sector, being the
dominant source of theoretical error in the short distance
contribution to $\Delta_{LR}$, should not be neglected.
\item
In the case $x\ll 1$, which is relevant for the charm contribution,
the function $Y$ has a very special structure. Expanding the
renormalization group result $Y_{NL}$ to first order in $\alpha_s$
one finds (here $x=m^2_c/M^2_W$)
\b\l{ynl1}
Y_{NL}\doteq {x\o 2}+{\alpha_s\o 4\pi}x \ln^2x  \e
We observe that the leading logarithms $\sim x\ln x$, present in
the Z-penguin- and the box part, have canceled in $Y_{NL}$, leaving
the subleading term $x/2$ as the only contribution in the limit
$\alpha_s=0$. On the other hand QCD effects generate an
$\alpha_s x\ln^2 x$ "correction", which is of the order
${\cal O}(x\ln x)$, hence a leading logarithmic term! As a first
consequence the charm function $Y$ is enhanced considerably
(by a factor of $\sim 2.5$) through strong interaction corrections,
compared to the non-QCD result. (This feature is in a sense similar
to the case of the rare decay $b\to s\gamma$.) A second point is
that the $x/2$ term, though formally subleading, is important
numerically. Working within LLA one is then faced with the problem
of how to deal with this term since it should strictly speaking
be omitted in this approximation. Let us illustrate this issue
in terms of the $\bar\varrho$ determination in our above example.
We find $\bar\varrho=-0.12$ if we use the LLA formulae
(with $\mu_c=m_c$, $\mu_t=m_t$) and simply add the $x/2$ piece.
By contrast, omitting this term and using the strict leading log result
we obtain $\bar\varrho=-0.20$. The scale ambiguities are very
similar in both cases, roughly three times as big as in the
next-to-leading order discussed above.
For definiteness we have included the $x/2$ part to obtain \r{rhor0}.
\\
The problem of the $x/2$ term is naturally removed in the
next-to-leading logarithmic approximation ($Y_{NL}$) where this
contribution is consistently taken into account.
\end{itemize}
Finally we give the standard model expectation for $\Delta_{LR}$,
based on the short distance contribution in \r{2}, for the
Wolfenstein parameter $\varrho$ in the range
$-0.25\leq\varrho\leq 0.25$, $V_{cb}=0.040\pm 0.004$ and
$m_t=(170\pm 20)GeV$. Including the uncertainties due to
$m_c$, $\Lambda_{\overline{MS}}$, $\mu_c$ and $\mu_t$ and
imposing the cut $-0.5\leq\cos\theta\leq 1$, we find
\b\l{dlr1}
3.0\cdot 10^{-3}\leq |\Delta_{LR}|\leq 9.6\cdot 10^{-3}   \e
employing next-to-leading order formulae.
Anticipating improvements in $V_{cb}$, $m_t$ and $\varrho$ we also
consider a future scenario in which
$\varrho=0.00\pm 0.02$, $V_{cb}=0.040\pm 0.001$ and
$m_t=(170\pm 5)GeV$. The very precise determination of $\varrho$
used here should be achieved through measuring CP asymmetries
in B decays in the LHC era \cite{BU}. Then \r{dlr1} reduces to
\b\l{dlr2}
4.8\cdot 10^{-3}\leq |\Delta_{LR}|\leq 6.6\cdot 10^{-3}   \e
In both of the scenarios the lower (upper) limit for $\Delta_{LR}$
would be smaller by $0.6\cdot 10^{-3}$ ($1.3\cdot 10^{-3}$) if the
charm contribution was omitted.

In this letter we have generalized the short distance calculation
of the muon polarization asymmetry $\Delta_{LR}$ in the decay
\kpmm to next-to-leading order in QCD. We furthermore discussed the
theoretical uncertainties involved in this analysis. We have
demonstrated that the complete next-to-leading order calculation
achieves a reduction of the rather large scale ambiguities
in leading order by a factor of $\sim 3$ and is necessary to
provide a satisfactory treatment of $\Delta_{LR}$. This is
particularly important since long distance contributions to
$\Delta_{LR}$ seem to be small, though perhaps not fully
negligible \cite{SW2}.
\\
In any case a measurement of $\Delta_{LR}$ would yield a very
interesting and useful piece of information on short
distance flavordynamics and the unitarity triangle which is
worth pursuing in future experiments.

\vfill\eject

\end{document}